\documentclass[aps,prd,reprint,preprintnumbers,groupedaddress,nofootinbib]{revtex4-2}

\usepackage{caption} 
\usepackage[T1]{fontenc}
\usepackage[latin9]{inputenc}
\usepackage[active]{srcltx}
\usepackage{graphicx}
\usepackage{amsmath,amsthm,amssymb,braket,cancel,enumerate,float,graphicx,mathtools,tabu,tikz,subcaption}
\usepackage[hidelinks]{hyperref}
\usepackage{cleveref}
\usepackage[export]{adjustbox}
\usepackage{booktabs}
\usepackage{multirow}
\usepackage{makecell}
\usepackage{tikz}
\usepackage{tikz-feynman}

\makeatletter

\renewcommand\labelenumi{(\theenumi)}

\newcommand{\abs}[1]{\left|{#1}\right|} 

\newcommand{\met}{E_{\text{T}}^{\mathrm{miss}}}

\@ifundefined{textcolor}{}
{%
 \definecolor{BLACK}{gray}{0}
 \definecolor{WHITE}{gray}{1}
 \definecolor{RED}{rgb}{1,0,0}
 \definecolor{GREEN}{rgb}{0,1,0}
 \definecolor{BLUE}{rgb}{0,0,1}
 \definecolor{CYAN}{cmyk}{1,0,0,0}
 \definecolor{MAGENTA}{cmyk}{0,1,0,0}
 \definecolor{YELLOW}{cmyk}{0,0,1,0}
}

\makeatother

\usepackage{babel}

\begin{document}

\begin{flushright}
 {\tt
CTPU-PTC-23-41  \\
}
\end{flushright}

\title{Taking aim at the wino-higgsino plane with the LHC}

\author{Linda M. Carpenter}
\email{lmc@physics.osu.edu}
\affiliation{Department of Physics, The Ohio State University\\ 191 W. Woodruff Ave., Columbus, OH 43210, U.S.A.}
\author{Humberto Gilmer}
\email{humberto_gilmer@brown.edu}
\affiliation{Brown Center for Theoretical Physics \\
Department of Physics, Brown University\\ Providence, Rhode Island 02912, U.S.A.}

\author{Junichiro Kawamura}
\email{junkmura13@gmail.com}
\affiliation{Particle Theory  and Cosmology Group, Center for Theoretical Physics
of the Universe\\ Institute for Basic Science (IBS), Daejeon, 34126, Korea}

\author{Taylor Murphy}
\email{murphy@lpthe.jussieu.fr}
\affiliation{Laboratoire de Physique Th\'{e}orique et Hautes \'{E}nergies (LPTHE), UMR 7589\\ Sorbonne Universit\'{e} \& CNRS\\ 4 place Jussieu, 75252 Paris Cedex 05, France}

\begin{abstract}
In this work we explore multiple search strategies for higgsinos and mixed higgsino-wino states in the MSSM and project the results onto the $(\mu,M_2)$ plane. Assuming associated production of higgsino-like pairs with a $W/Z$ boson, we develop a search in a channel characterized by a hadronically tagged vector boson accompanied by missing energy. We use as our template an ATLAS search for dark matter produced in association with a hadronically decaying vector boson, but upgrade the search by implementing a joint-likelihood analysis, binning the missing transverse energy distribution, which greatly improves the search sensitivity. For higgsino-like states (more than $96\%$ admixture) we find sensitivity to masses up to 550\,GeV. For well-mixed higgsino-wino states ($70$--$30\%$ higgsino) we still find sensitivities above 300\,GeV. Using this newly proposed search, we draw a phenomenological map of the wino-higgsino parameter space, recasting several complementary searches for disappearing tracks, soft leptons, trileptons, and hadronic diboson events in order to predict LHC coverage of the $(\mu,M_2)$ mass plane at integrated luminosities of up to $3\,\text{ab}^{-1}$. Altogether, the full run of the HL-LHC can exclude much of the ``natural'' ($\mu,M_2 < 500\,\text{GeV}$) wino-higgsino parameter space.
\end{abstract}

\maketitle

\section{Introduction}\label{s1}

The Large Hadron Collider (LHC) has probed deeply into the low-mass parameter space of supersymmetry (SUSY). Gluinos are bounded below 2\,TeV and squark mass bounds are not much behind at 1.6\,TeV \cite{ATLAS-CONF-2019-040, ATLAS-CONF-2020-002, Sirunyan:2019xwh, Sirunyan:2019ctn}. Despite progress in searches for color-charged states, however, bounds on weakly interacting SUSY particles are not strong. In particular, bounds on electroweakinos with compressed mass spectra (degenerate or nearly degenerate) leave many unconstrained regions of parameter space. The situation is most striking for higgsino-like states and higgsino-wino admixtures. These states are important to bound as these particles appear as LSPs or NLSPs in a range of viable scenarios in the Minimal Supersymmetric Standard Model (MSSM) including general gauge mediation \cite{Carpenter:2008he, Carpenter:2017xru, Rajaraman:2009ga, Carpenter:2008wi}, anomaly mediation \cite{Giudice:1998xp,Randall:1998uk}, scenarios featuring non-universal gaugino masses \cite{Abe:2007kf,Abe:2012xm,Antusch:2012gv,Baer:2015tva}, mirage mediation \cite{Choi:2005hd,Kitano:2005wc,Choi:2006xb,Everett:2008qy,Abe:2014kla} and the higgsino world \cite{Baer:2011ec}.

The choice of collider search for electroweakinos depends on the region of parameter space to be probed, and especially on the mass splitting between the lightest neutralino $\tilde{\chi}^0_1$ and lightest chargino(s) $\tilde{\chi}^{\pm}_1$ or the next-lightest neutralino $\tilde{\chi}^0_2$. This splitting is in turn heavily dependent on the gauge-eigenstate composition of the light electroweakinos, which influences the nature of the search conducted. We suppose throughout this work that the bino with mass $M_1$ is decoupled and focus on the wino-higgsino mass plane. In this regime, for extremely wino-like particles, small mass splitting between charged and neutral states means there is sure to be a long-lived charged particle in events, motivating searches for long-lived charged tracks \cite{ATLAS:2017oal,CMS:2018rea,CMS:2020atg,Fukuda:2017jmk} or soft displaced tracks \cite{Fukuda:2019kbp}. For states with larger mass splittings between particles, on the other hand, searches with soft leptons may apply \cite{CMS:2015epp,ATLAS:2017vat,CMS:2018kag,Schwaller:2013baa, Han:2014kaa,Baer:2014kya,Baer:2021srt}. But there is a large gap in this search space where electroweakinos that are predominantly higgsino-like --- or a well-tempered mixture of wino-higgsino content --- where there are no long-lived charged tracks, and small splittings ensure mass degenerate states are more likely to appear as invisible particles. This window covers a large region in the $(\mu,M_2)$ mass plane of fundamental parameters. In this case a new search strategy is needed to improve coverage of the electroweakino parameter space. 

For intermediate mass splittings between $\tilde{\chi}^0_1$ and $\tilde{\chi}^{\pm}_1$ or $\tilde{\chi}^0_2$, the chargino and second-to-lightest neutralino states may be produced and decay with products so soft as to be considered missing energy by the search. In this case it is possible to trigger on the decay of a single heavy vector boson produced in association with electroweakino pairs. We choose to search for the heavy boson(s) in a hadronically tagged channel, continuing a line of inquiry begun in \cite{Carpenter:2020fnh,Carpenter:2021jbd} targeted at rare and hard-to-constrain SUSY signals.\footnote{A dijet signal may also be of interest \cite{Buanes:2022wgm}.} Our hadronic mono-boson analysis is based on a search by the ATLAS Collaboration for jets accompanied by missing transverse energy ($E_{\text{T}}^{\text{miss}}$) \cite{Aaboud:2018xdl}, which we extend by performing a joint-likelihood analysis using the $E_{\text{T}}^{\text{miss}}$ distributions. Our strategy significantly improves the sensitivity of the original ATLAS search to electroweakino pair production and allows us to close an existing hole in the electroweakino parameter space not covered by other searches.

The aim of this work is to project the bounds from this mono-boson channel to the $(\mu, M_2)$ plane, and compare it with the other channels at the LHC today and in the future. To this end, alongside our own analysis, we reinterpret four existing analyses that are expected to be sensitive to wino-higgsino LSP scenarios. These searches are in channels characterized by disappearing tracks \cite{CMS:2020atg}, soft leptons \cite{CMS:2018kag}, three leptons accompanied by missing transverse energy \cite{ATLAS:2021moa}, and two hadronically decaying vector bosons with missing energy \cite{ATLAS:2021yqv}. We present a phenomenological map of the wino-higgsino mass plane detailing which searches are most sensitive at present and for the projected $3\,\text{ab}^{-1}$ HL-LHC run. We find complementarity between the searches, with the mono-boson search able to cover a sizable region of parameter space. We expect the full run of the HL-LHC to probe or exclude almost all of the ``natural'' (small-$\mu$) wino-higgsino parameter space.

This paper proceeds as follows. In \hyperref[s2]{Section II} we review the masses and splitting of electroweakinos in the MSSM. \hyperref[s3]{Section III} concerns the electroweakino parameter space and searches that cover its various regions. In \hyperref[s4]{Section IV} we describe our hadronic mono-boson search strategy. \hyperref[s5]{Section V} presents results of a sensitivity search for the HL-LHC. \hyperref[s6]{Section VI} concludes. 

\section{Wino-higgsino spectra in the MSSM}\label{s2}

We begin with a brief review of the spectrum of the electroweakinos relevant to our study. 
Concretely, since we are interested in higgsinos, higgsino-wino admixtures, and winos, we focus on the hierarchy $\mu,M_2\ll M_1$. A higgsino state corresponds to $\mu < M_2$, a well-mixed state to $\mu\sim M_2$, and a wino state to $\mu > M_2$. In the higgsino limit, $\mu < M_2$, the eigenvalues of $\mathbf{M}_{\tilde{\chi}^0}$ may be approximated in order of increasing mass as \cite{PhysRevD.37.2515}
\begin{align}
\nonumber    
m_{\tilde{\chi}^0_{1,2}} &= \begin{multlined}[t][3cm]\mu + m_Z^2\,\frac{(1 \mp \sin2\beta)(\mu + M_2\text{s}^2_{\text{w}} + M_1\text{c}^2_{\text{w}})}{2\left(\mu \pm M_2\right)\left(\mu \pm M_1\right)},\end{multlined}
    \\
m_{\tilde{\chi}^0_{3,4}} &= M_{2,1} - m_W^2\,\frac{M_{2,1} + \mu\sin2\beta}{\mu^2 - M^2_{2,1}}
    \label{eq:neutralino_masses}
\end{align}
for $\mu>0$, a choice we adopt in this work. 
In the light wino limit (still with $M_1$ decoupled), the mass ordering in \eqref{eq:neutralino_masses} changes from $\{1,2,3,4\}$ to $\{3,1,2,4\}$.
\begin{figure}
    \centering
    \includegraphics[width=0.44\textwidth]{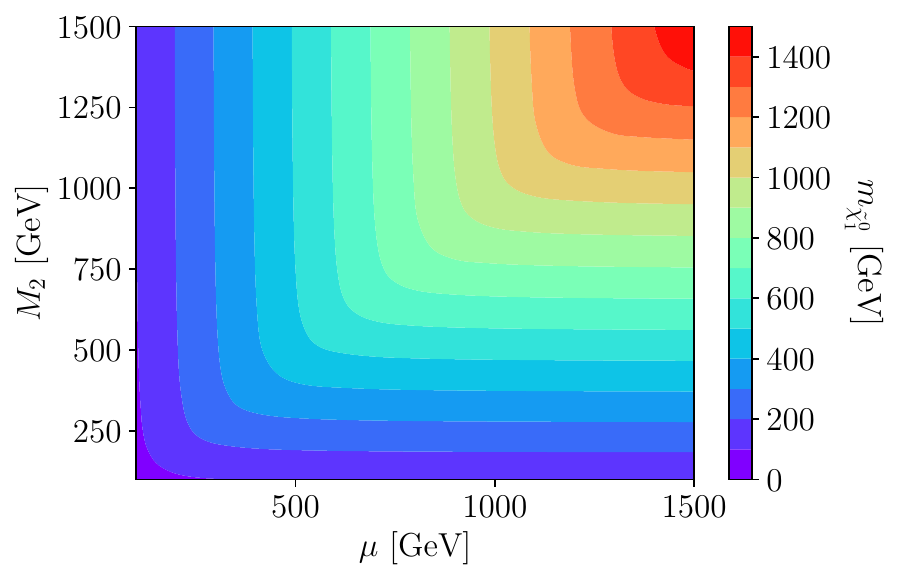}
    \caption{\label{fig:N1_mass} Contour plot of the lightest neutralino mass $m_{\chi_1^0}$ in the  $(\mu,M_2)$ plane.}
\end{figure}Similarly, for charginos in the higgsino limit, the mass eigenvalues are approximately
\begin{align}
 \nonumber   m_{\tilde{\chi}^{\pm}_1} &= \mu + m_W^2\, \frac{\mu + M_2\sin2\beta}{\mu^2 - M^2_2}
    \\
\text{and}\ \ \    m_{\tilde{\chi}^{\pm}_2} &= M_2 - m_W^2\,\frac{\mu + M_2\sin2\beta}{\mu^2 - M^2_2}
    \label{eq:chargino_masses}
\end{align}
with hierarchy flipped in the wino limit. Note that in the deep higgsino region, $m_{\tilde{\chi}^0_1}\sim m_{\tilde{\chi}^{\pm}_1}$; as we will see, the mass difference between the lightest chargino and lightest neutralino, given by
\begin{align}
\Delta m = m_{\tilde{\chi}^{\pm}_1} - m_{\tilde{\chi}^0_1}
  \simeq \frac{m_W^2}{2 M_2}\frac{(1-\tan\beta)^2}{1+\tan^2\beta},
\end{align}
will play an important role in our search strategy.

In this work we are concerned with states that have a naturally small mass difference between charginos and neutralinos, as this is the most technically challenging part of the electroweakino parameter space to probe experimentally. Both wino-like and higgsino-like neutralinos feature some naturally small mass differences, with the former scenario exhibiting nearly degenerate $\{\tilde{\chi}^0_1,\tilde{\chi}^{\pm}_1\}$ and the latter providing nearly triply degenerate $\{\tilde{\chi}^0_1,\tilde{\chi}^{\pm}_1,\tilde{\chi}^0_2\}$. Therefore in this work we consider scenarios where $M_1$ is very large, leaving us with wino- or higgsino-like (light) neutralino parameter space. 
For the purposes of this work, we fix $\tan\beta = 10$, which is a common choice but not particularly important for the electroweakino splittings: for instance, raising $\tan \beta$ to values as large as 100 shifts the physical masses by $\mathcal{O}(1)\,\text{GeV}$ but has negligible effects on the mass differences. This choice leaves us with only $\mu$ and $M_2$ as adjustable parameters. 

The importance of small mass splittings in this analysis requires us to go beyond leading-order calculations of the electroweakino masses, since one-loop corrections to light masses can approach $10\%$ of the leading-order results \cite{PIERCE19973}. We employ \textsc{SPheno} version 4.0.5 \cite{Porod:2003um,Porod:2011nf,Staub:2017jnp} to compute the mass spectra (and mixing matrices) for a large number of points in the $(\mu,M_2)$ plane. In Figure \ref{fig:N1_mass} we show the mass of the lightest neutralino $\tilde{\chi}_1^0$ in this plane for $\tan\beta=10$ and $M_1= 5\,\text{TeV}$.  

\begin{figure}
\centering
    \begin{subfigure}{0.44\textwidth}
        \includegraphics[width=\textwidth]{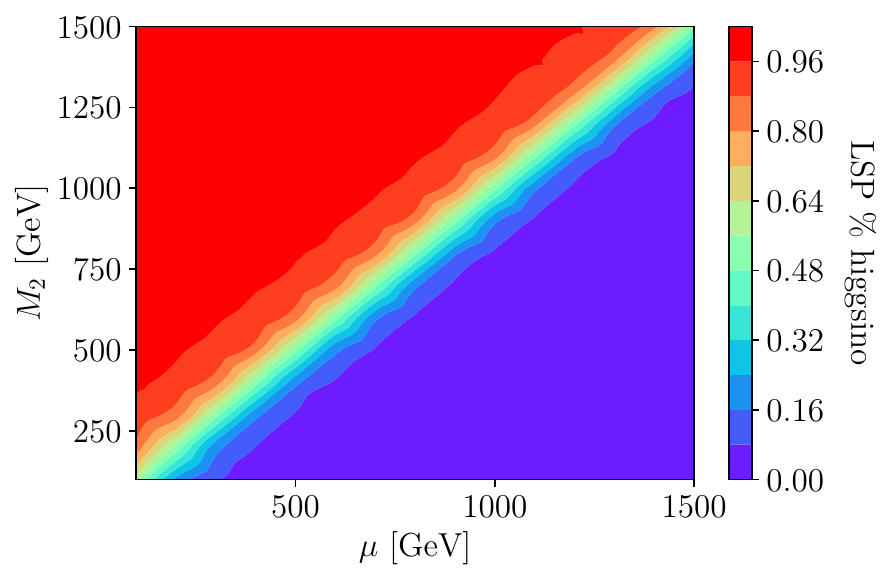}
        \caption{\label{fig:mixing_plane}}
    \end{subfigure}
\hfill
    \begin{subfigure}{0.44\textwidth}
        \includegraphics[width=\textwidth]{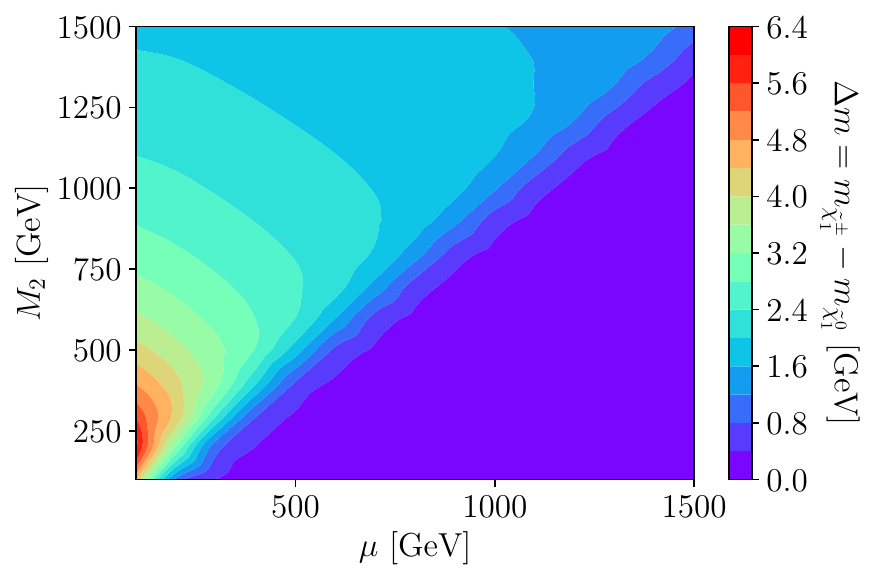}
        \caption{\label{fig:N1C1_mass_difference_plane}}
    \end{subfigure}
    \caption{Contour plots of (\subref{fig:mixing_plane}) the higgsino content of the LSP and (\subref{fig:N1C1_mass_difference_plane}) the mass difference $\Delta m = m_{\tilde{\chi}^{\pm}_1} - m_{\tilde{\chi}^0_1}$.}
\end{figure}

Both the content of the lightest neutralino and the magnitude of the mass splitting between the lightest chargino and the lightest neutralino vary over the mass plane. In Figure \ref{fig:mixing_plane} we show the higgsino content of the lightest neutralino over the $(\mu,M_2)$ plane for our benchmark values of $M_1$ and $\tan\beta$. In Figure \ref{fig:N1C1_mass_difference_plane} we show the mass difference $\Delta m$ between the lightest chargino and the lightest neutralino in the same plane.  We see that in the wino-like region with $\mu \gg M_2$, the mass splitting between chargino and wino is very small, less than 1\,GeV. In the higgsino-like region, the mass splitting is still small on an absolute scale but varies from one to a few GeV. There is also a well-mixed region in which the higgsino content varies from 30-70\%.

\begin{figure}
    \centering
    \includegraphics[width=0.485\textwidth]{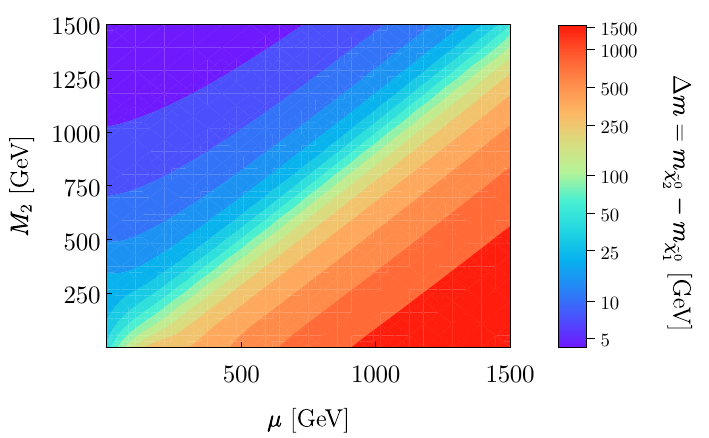}
    \caption{\label{fig:N1N2_difference} Contour plot of the mass difference between the lightest and next-lightest neutralino, $\Delta m = m_{\tilde{\chi}_2^0} - m_{\tilde{\chi}_1^0}$}.
\end{figure}

The mass of the second-lightest neutralino $\tilde{\chi}_2^0$ also varies dramatically over the parameter space. Figure \ref{fig:N1N2_difference} shows the mass splitting between the lightest and second-lightest neutralinos; that is, $m_{\tilde{\chi}_2^0}-m_{\tilde{\chi}_1^0}$, in the $(\mu,M_2)$ plane. We see that for higgsino-like states the mass splitting is small. As we transition across the mass plane to well-mixed and wino-like states the mass splitting increases to $\mathcal{O}(100)\,\text{GeV}$ and more.  As we will later see, the production and decay of $\tilde{\chi}_2^0$ will also greatly influence the neutralino searches.

\section{Probing $\boldsymbol{(\mu,M_2)}$ with multiple search strategies}
\label{s3}

We now consider the LHC phenomenology of the light electroweakinos in our parameter space.  While lightest neutralinos $\tilde{\chi}_1^0$ invariably appear in the detector as invisible particles, the charginos may decay visibly or invisibly. For small mass differences, the chargino decay proceeds through an off-shell $W$, $\tilde{\chi}^{\pm}_1\rightarrow\tilde{\chi}_1^0+W^{\pm *}$. Exactly how these decays appear in the detector, hence how best to probe the charginos experimentally, depends sensitively on the mass splitting. 
We identify three $\Delta m$ regimes:
\begin{enumerate}
\renewcommand{\labelenumi}{(\Alph{enumi})}
    \item {Long-lived charginos, $\Delta m \lesssim 1$ GeV;} 
    \item {Charginos making soft leptons, $\Delta m \gtrsim 4$ GeV; and} 
    \item {Invisible charginos, $1 \lesssim \Delta m \lesssim 4$ GeV;}
\end{enumerate}
each best suited to a unique search strategy. In Figure \ref{fig:param_space_cartoon} we show the parameter space plane for higgsino- and wino-like LSPs with $M_1 = 5\,\text{TeV}$. 
In this plane we demarcate the chargino-neutralino mass splitting in order to sketch the parameter space best suited to the search strategies detailed below. We have also marked in orange the threshold above which the LSP is higgsino-like, which we define as greater than 96\% higgsino content.

In the following discussion, we overview the search strategies in these three regimes. 
The details of the mono-boson search, which is our main result, will be explained in the next section. 

\subsection{Nearly degenerate charginos: the long-lived particle region}\label{s3.1}

\begin{figure}
    \centering
    \includegraphics[width=0.45\textwidth]{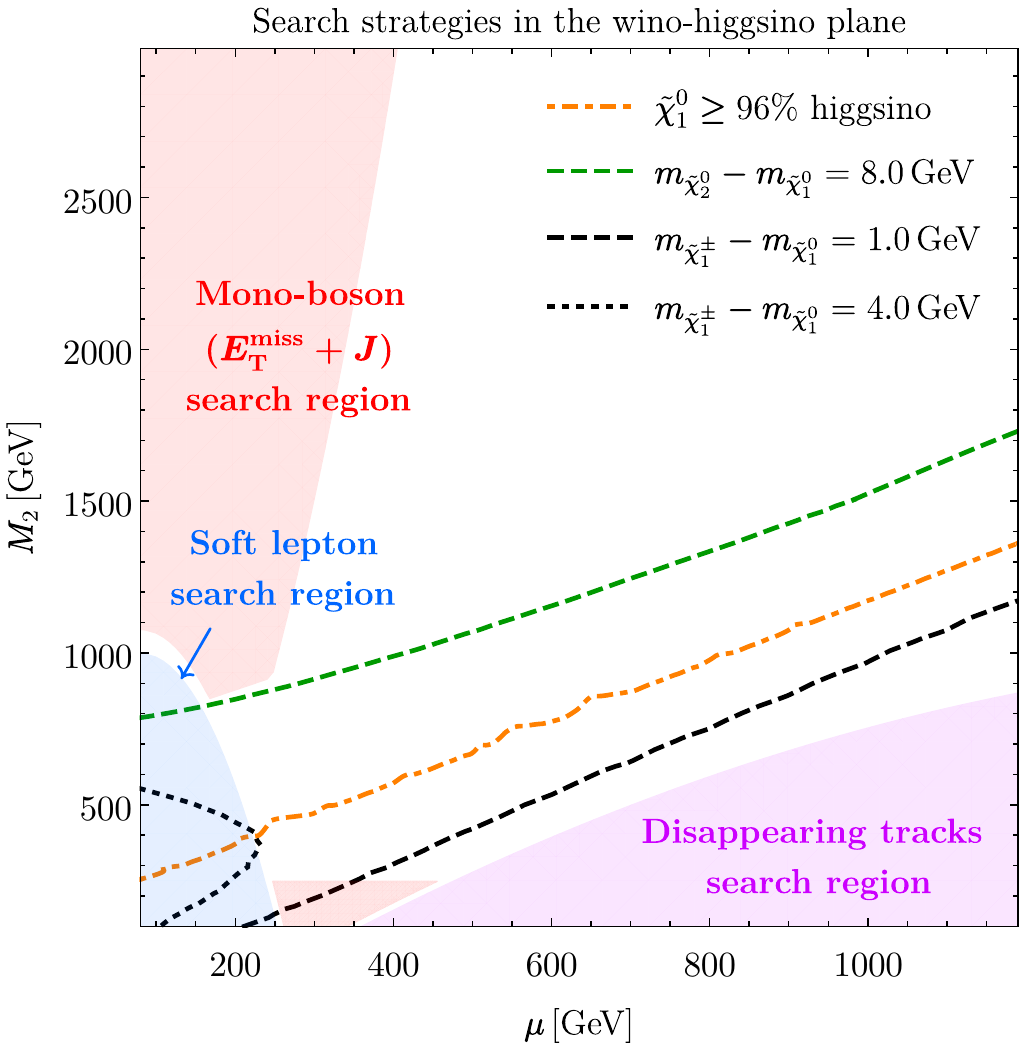}
    \caption{
    Search strategies in the $(\mu,M_2)$ plane based on mass difference. Also shown is the wino vs. higgsino content of the lightest neutralino. Recall from Figures \ref{fig:N1N2_difference} and \ref{fig:N1C1_mass_difference_plane} that $m_{\tilde{\chi}^0_2}-m_{\tilde{\chi}^0_1}$ increases with $\mu$ but $m_{\tilde{\chi}^{\pm}_1}- m_{\tilde{\chi}^0_1}$ does the opposite.
    \label{fig:param_space_cartoon}}
\end{figure}

For the smallest mass splittings, the decay products are very soft, so detecting production of pairs such as $\tilde{\chi}^{\pm}_1 \tilde{\chi}^{\mp}_1$, $\tilde{\chi}_i^0\tilde{\chi}^{\pm}_1$, $i,j=1,2$, cannot rely on hard leptons or jets. We see in Figure \ref{fig:param_space_cartoon} that under the black dashed line the mass splitting between the lightest chargino and lightest neutralino is under 1\,GeV. There is a portion of this wino-like LSP parameter space where the lightest chargino lives long enough to produce a track in a detector, and so searches for long-lived tracks are expected to give the best mass bounds on LSPs.

An applicable search of this type was performed by the CMS Collaboration using $\mathcal{L} = 101\,\text{fb}^{-1}$ of Run 2 data and published as CMS-EXO-19-010 \cite{CMS:2020atg}. This search targets long-lived charged particles, like our charginos, exhibiting ``disappearing'' tracks that leave the interaction region but do not extend to the outermost region of the tracking detector. A track is defined to disappear if it has at least three missing outer hits in the tracker and if the total calorimeter energy within $\Delta R = 0.5$ of the track is less than $10\,\text{GeV}$. This search applies to charged particles with lifetimes in the range $\tau \in [0.3,333]\,\text{ns}$ (the low end of this range is self explanatory; the high end is a practical limit past which charged particles live \emph{too} long and their tracks do not disappear before the edge of the tracker.

In the absence of an excess, CMS imposed limits on chargino production in a few supersymmetric scenarios, including models with higgsino- and wino-like electroweakinos, the latter of which is appropriate for our analysis. This search has moreover been implemented within the \textsc{MadAnalysis\,5} (MA5) framework \cite{Conte:2012fm,Conte:2014zja,Dumont:2014tja} and made available on the MA5 Public Analysis Database (PAD) \cite{Conte:2018vmg,Araz:2020lnp}. In order to reinterpret the CMS results within our parameter space, we use \textsc{MadGraph5\texttt{\textunderscore}aMC@NLO} (\textsc{MG5\texttt{\textunderscore}aMC}) version 3.1.0 \cite{Alwall:2014hca} to produce a number of electroweakino pair-production samples in the pink region of the $(\mu,M_2)$ plane depicted in Figure \ref{fig:param_space_cartoon}. These samples need to be relatively large, each containing $2.5 \times 10^5$ events, to maintain statistical control given the very low efficiencies characteristic of this search \cite{Araz:2021akd}. We simulate showering and hadronization with {\sc Pythia\,8} version 8.245 \cite{Sjostrand:2014zea}, which also handles the decays of the electroweakinos. We extract the electroweakino decay widths and branching fractions from \textsc{SPheno} version 4.0.5, mentioned in \hyperref[s2]{Section II} as the generator of our mass spectra. The widths, like the masses, are accurate to one-loop order, which is crucial for \emph{e.g.} $\tilde{\chi}^{\pm}_1$ decays to pions \cite{Goodsell:2020lpx}. To set the normalization of the samples, we use \textsc{Resummino} version 3.1.2 \cite{Fiaschi:2023tkq} to compute the total cross sections of lightest chargino and/or LSP pair production for $\sqrt{s}=13\,\text{TeV}$ and $\sqrt{s}=14\,\text{TeV}$ at approximate next-to-next-to-leading-order accuracy in the strong coupling with threshold resummation at next-to-next-to-leading logarithmic accuracy (aNNLO + NNLL). These showered and hadronized event samples are then passed to MA5 version 1.9.60, which uses the Simplified Fast Detector Simulation (SFS) module \cite{Araz:2020lnp} to simulate the response of the CMS detector and calls \textsc{FastJet} version version 3.3.3 for object reconstruction \cite{FJ}. When MA5 is provided with the signal cross sections, it computes not only the upper limit at 95\% confidence level (C.L.) \cite{Read:2002cls} on the cross section of \emph{any} bSM signal, given the efficiencies returned in each signal region of the analysis, but also the signal confidence level $\text{CL}_s$ of the particular signal given by the user, such that the signal is excluded if $\text{CL}_s = 0.05$. The recasting capabilities of MA5 moreover include higher-luminosity estimates, which rescale the signal and background yields linearly with luminosity and rescale the yield uncertainties according to the user's preferences \cite{Araz:2019otb}. We use this module to provide sensitivity estimates for the $\mathcal{L}=3\,\text{ab}^{-1}$ run of the HL-LHC. For this exercise, we use cross sections computed at a center-of-mass energy of $\sqrt{s}=14\,\text{TeV}$, but we use the same $\sqrt{s}=13\,\text{TeV}$ event samples due to the significant computational resources required to produce the samples discussed here. The results of this reinterpretation, along with those described below, are discussed in \hyperref[s5]{Section V}.

\subsection{Locally maximal chargino splitting: the soft lepton region}\label{s3.2}

In the region above the black dashed line the chargino-LSP splitting exceeds 1\,GeV. Both higgsino-like and mixed wino-higgsino regions in this parameter space have small $\Delta m$, but in the region where both $\mu$ and $M_2$ are small, the splitting attains a local maximum. For our benchmark with $\tan\beta = 10$, the maximum mass difference is $\Delta m \sim 6\,\text{GeV}$. In Figure \ref{fig:param_space_cartoon}, we have marked the $\Delta m = 4\,\text{GeV}$ threshold with the black dotted line. In the region enclosed to the left of this curve, there may be soft but detectable leptons from chargino decay. Meanwhile, adjoining the same region in parameter space, the  mass splitting $m_{\chi^0_2}-m_{\chi^0_1}$ between the two lightest neutralinos becomes appreciable. On the plot we have marked with a dashed green line the region under which this mass splitting is greater than 8\,GeV. Roughly between the line demarcating the higgsino-like LSP region and this green dashed line, we expect small but relevant lepton momentum from decays through off-shell $W/Z$ bosons. In this space, the electroweakino spectrum is still ``compressed'', but leptons resulting from $\tilde{\chi}^{0}_2$ decays --- while quite soft --- have enough momentum in principle to be detected at LHC. We therefore expect searches for events with soft leptons to impose non-trivial limits in this region.

One such soft-lepton search was carried out by CMS using $\mathcal{L} = 35.9\,\text{fb}^{-1}$ of Run 2 data and was published as CMS-SUS-16-048 \cite{CMS:2018kag}. This search notably requires two leptons with transverse momentum $p_{\text{T}} < 30\,\text{GeV}$ and, finding no excesses, was used to constrain several benchmark supersymmetric models with electroweakino mass splitting of $\mathcal{O}(1\text{--}10)\,\text{GeV}$. One of the constrained scenarios features compressed higgsino-like electroweakinos, but \emph{a priori} this analysis could be sensitive to well-mixed species. We therefore make use of the public implementation of this analysis in the \textsc{MadAnalysis\,5} PAD, according to a workflow similar to that discussed above for the disappearing-tracks analysis, to reinterpret the soft-lepton search in our parameter space and to compute HL-LHC sensitivity estimates.

Before we move on, it is worth noting that searches for electroweakinos in final states with three leptons also analyze $\tilde{\chi}^{\pm}_1$ and $\tilde{\chi}^{0}_2$ production, with leptons resulting from the decay of these states to the LSP. Current trilepton analyses are capable of sensitivity in regions where the $\tilde{\chi}^{0}_2$-$\tilde{\chi}^{0}_1$ mass splitting is as little as a few GeV, which overlaps with our soft-dilepton region \cite{ATLAS:2021moa}. Therefore, as we demonstrate in \hyperref[s5]{Section V}, trilepton searches are capable of imposing some limits in this area.

\subsection{The invisible chargino region\label{s3.3}}

In the case of a mass splitting just large enough for the charginos to decay promptly, but not large enough to produce hard particles to trigger on, both charginos and neutralinos are effectively invisible. In this parameter space we must rely on an alternate strategy: the production of light electroweakinos --- recorded as missing transverse energy $E_{\text{T}}^{\text{miss}}$ --- along with an on-shell vector boson, $pp\rightarrow \tilde{\chi}\tilde{\chi} + W/Z$. Here, the on-shell boson decays hadronically and may be tagged. This search is best suited for higgsino-like or higgsino-wino-like LSPs (those outside of the deep wino region) for the following reasons.

\begin{itemize}
    \item In the higgsino-like region, there are three nearly degenerate electroweakinos $\{\tilde{\chi}^{0}_1,\tilde{\chi}^{0}_2,\tilde{\chi}^{\pm}_1\}$. Due to the softness of their decay products, all three of these states may appear as invisible particles in the detector, and any pair of these particles may be produced in association with an on-shell $W/Z$ boson.

    \item As we can see from Figures \ref{fig:N1C1_mass_difference_plane} and \ref{fig:N1N2_difference}, as we move into the more well-mixed region the chargino remains mass degenerate with the lightest neutralino, but $\tilde{\chi}_2^0$ achieves greater mass splitting. As the mass splitting grows to $\mathcal{O}(10)\,\text{GeV}$, $\tilde{\chi}_2^0$ is no longer an invisible particle and therefore the total production cross section for our invisible process plus a gauge boson is apparently diminished.
    
    \item But farther toward the wino region, where the $\tilde{\chi}^0_2$ splitting exceeds the mass of the $W$ or $Z$, $\tilde{\chi}_2^0$ may decay to  $\tilde{\chi}_1^0$ or $\tilde{\chi}^{\pm}_1$ through an on-shell vector boson. This gives us the process $pp\rightarrow \tilde{\chi}_2^0 \tilde{\chi}^{\pm,0}_1 \rightarrow \tilde{\chi}^{\pm,0}_1\tilde{\chi}^{\pm,0}_1 +V$, with a hard vector boson radiated as a decay product in the final state. The jet(s) produced by hadronically decaying vector bosons should be correspondingly hard.
\end{itemize}
The red regions in Figure \ref{fig:param_space_cartoon} (``$E_{\text{T}}^{\text{miss}} + J$'') are therefore roughly where we expect a hadronic mono-boson search to set the best limits. In the well-mixed region, the mono-boson analysis should complement not only the CMS soft-lepton search detailed above but also conventional searches in channels with more than one hadronic vector boson \cite{ATLAS:2021yqv} or with multiple leptons \cite{CMS:2017moi,ATLAS:2019lff,ATLAS:2019wgx,ATLAS:2021moa}. A quantitative comparison verifying this notion is available in \hyperref[s5]{Section V}.

We note here, in advance of our detailed discussion of the mono-boson analysis, that the most recent limits from conventional monojet searches have historically been weaker and have less coverage of the $(\mu,M_2)$ plane than those from this mono-boson analysis. For this discussion, we refer to monojet limits on direct pair production of electroweakinos, assuming that the squarks are sufficiently heavy that monojet limits on electroweakinos due to pair production of light squarks do not apply. The situation for light electroweakinos was discussed in \cite{Anandakrishnan:2014exa} with respect to the Run 1 ATLAS monojet search \cite{ATLAS:2012ky} and in \cite{Carpenter:2021jbd} for the most recent Run 2 ATLAS monojet search \cite{ATLAS:2021kxv}. However, in the time since \cite{Carpenter:2021jbd} was released, both this ATLAS analysis and its CMS counterpart, the monojet subanalysis in CMS-EXO-20-004 \cite{CMS:2021far}, have been implemented in \textsc{MadAnalysis\,5}, and moreover a thorough analysis of monojet constraints on higgsinos has been released very recently \cite{Agin:2023yoq}. While the ATLAS search remains weak, and monojet constraints on winos are expected to be superseded by disappearing-track limits, the limits derived from a combination of the CMS monojet signal regions are competitive with our mono-boson limits for $\mu \ll M_2$. We therefore discuss the interplay between mono-boson and monojet higgsino limits in greater detail in \hyperref[s5]{Section V}.

\section{Custom hadronic mono-$\boldsymbol{W/Z}$ ($\boldsymbol{E_{\textbf{T}}^{\textbf{miss}} + J}$) analysis}\label{s4}

Our mono-boson analysis upgrades an existing search by the ATLAS Collaboration \cite{Aaboud:2018xdl} based on a partial LHC Run 2 dataset with integrated luminosity $\mathcal{L} = 36.1\,\text{fb}^{-1}$. Mono-boson searches were originally conceived for fermionic dark matter models, \emph{e.g.} \cite{Lopez:2014qja}. This ATLAS search targets single on-shell hadronically-decaying vector bosons, produced in association with invisible particles. The typical event topology features significant missing energy along with either $\geq \! 1$ fat jet or $\geq\! 2$ narrow jets. Here we discuss in greater detail the signals probed by this analysis before reviewing the ATLAS selections and detailing our enhanced analysis.

\subsection{Compressed electroweakino pair $\boldsymbol{+}$ hadronic $\boldsymbol{W/Z}$ production at LHC}
\label{s4.1}

\begin{figure}
\centering
\includegraphics[scale=1]{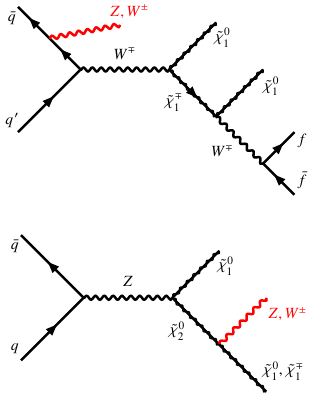}
    \caption{\label{fig:channel_diagrams} Representative parton-level diagrams for some channels considered in this work.}
\end{figure}

For this search, we consider hadronic collider processes of the form $pp\rightarrow \tilde{\chi}\tilde{\chi}+V$, where $\tilde{\chi} = \{\tilde{\chi}^0_1, \tilde{\chi}^0_2, \tilde{\chi}^{\pm}_1\}$ and where $V = \{W^{\pm}, Z\}$. Figure \ref{fig:channel_diagrams} shows schematic diagrams of the relevant processes, which we enumerated in \hyperref[s3.3]{Section III}.
In such processes, the momenta of the visible decay products depend heavily on the hardness of the associated vector bosons, which in turn is dependent on the mass splitting between the LSP and the lightest chargino $\tilde{\chi}^{\pm}_1$ or second-lowest-mass neutralino $\tilde{\chi}^0_2$. As established above, the regions of interest are the pure higgsino region, where $\tilde{\chi}^0_1$ is $96\%$ higgsino or higher, and the well-tempered higgsino-wino region where $\tilde{\chi}^0_1$ is $30\text{--}70\%$ higgsino.

\begin{figure}[t]
\centering
\includegraphics[width=0.95\columnwidth]{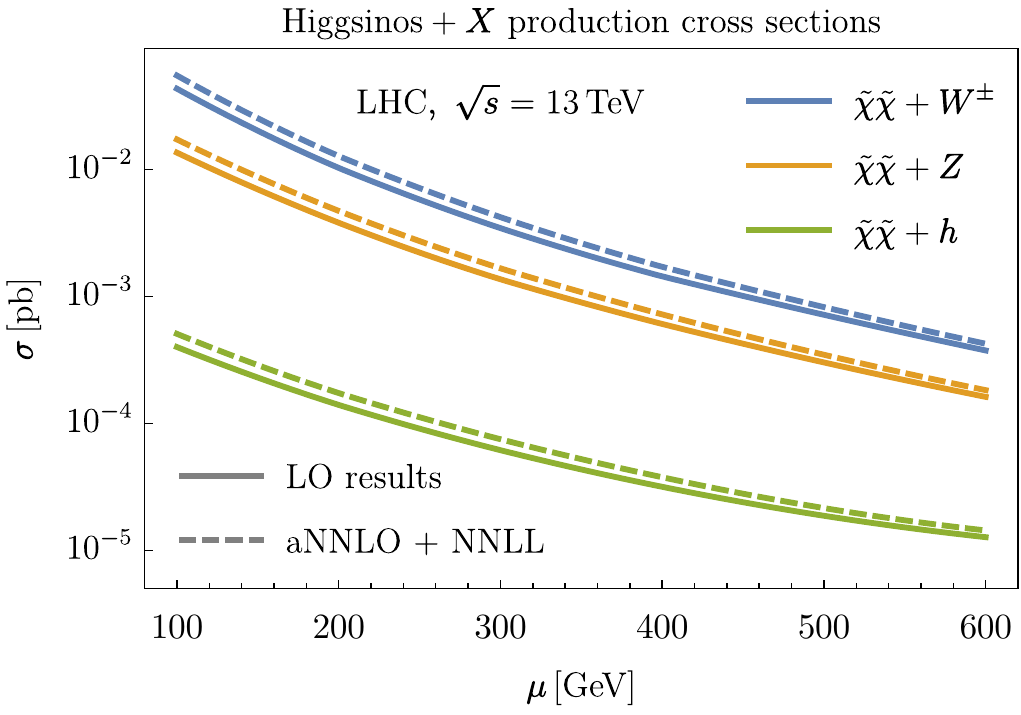}
    \caption{\label{fig:mono-boson-xsec}Cross sections of electroweakino pair + mono-boson processes for higgsino-like LSP ($M_1 = 5\,\text{TeV}$, $M_2 = 1\,\text{TeV}$). Here $\tilde{\chi}$ denotes $\{\tilde{\chi}^0_1,\tilde{\chi}^0_2,\tilde{\chi}^{\pm}_1\}$. Results are given at leading order (LO) and approximate next-to-next-to-leading order with next-to-next-to-leading logarithmic threshold resummation (aNNLO+NNLL).}
\end{figure}

In Figure \ref{fig:mono-boson-xsec} we have plotted typical production cross sections for pairs of light electroweak gauginos produced in association with $W/Z$ vector bosons in a slice of the higgsino-like parameter space. We specifically show the LHC production cross sections for $\sqrt{s}=13\,\text{TeV}$ as a function of $\mu$ with $M_2$ fixed at 1\,TeV. These results are given at LO and aNNLO + NNLL, as discussed in \hyperref[s3.1]{Section III}, and exhibit moderate $K$ factors in the range $K \sim (1.1,1.3)$, typical for such processes \cite{Debove:2010kf}. We see that generically production with an associated $W$ boson has the highest cross section. We have also included the cross section of associated production with a Higgs boson $h$ in order to demonstrate that its rate is much smaller than the mono-$V$ processes. 

\renewcommand{\arraystretch}{1.5}
\begin{table*}
 \centering
\begin{tabular}[t]{c||cccc|cc}\toprule\hline
                     & \multicolumn{4}{c}{Merged topology} & \multicolumn{2}{|c}{Resolved topology} \\   \hline\hline 
   $\met$      & \multicolumn{4}{c}{$>250$\,GeV }       & \multicolumn{2}{|c}{$>150$\,GeV} \\   
Jets, leptons & \multicolumn{4}{c}{$\ge 1J$, 0$\ell$}   & \multicolumn{2}{|c}{$\ge 2j$, 0$\ell$} \\   
 $b$-jets           & \multicolumn{4}{c}{no $b$-tagged jets outside of $J$}&\multicolumn{2}{|c}{$\le 2$ $b$-tagged small-$R$ jets} \\   
\hline 
                     & \multicolumn{6}{c}{$\Delta \phi (\vec{E}_{\text{T}}^\mathrm{miss},\, J\ \mathrm{or}\ jj ) > 2\pi/3$} \\   
  \multirow{2}{*}[-1.0ex]{\makecell{Multijet\\ suppression}}&\multicolumn{6}{c}{$\min_{i=1,2,3}\left[\Delta\phi(\vec{E}_{\text{T}}^\mathrm{\ miss},\, j_i)\right]>\pi/9$} \\ 
  & \multicolumn{6}{c}{$\abs{\vec{p}_{\text{T}}^\mathrm{\ miss}} > 30$\,GeV or $\ge 2$ $b$-jets} \\ 
           & \multicolumn{6}{c}{$\Delta \phi (\vec{E}_{\text{T}}^\mathrm{\ miss}, \vec{p}_{\text{T}}^\mathrm{\ miss} ) < \pi/2$}\\ 
\hline 
\multirow{2}{*}[-0.5ex]{\makecell{Signal\\ properties}}     & \multicolumn{4}{c}{} & \multicolumn{2}{|c}{$p_{\text{T}}^{j_1} > 45$\,GeV} \\   
 & \multicolumn{4}{c}{} & \multicolumn{2}{|c}{\ $\sum_i p_{\text{T}}^{j_i} > 120$ ($150$)\,GeV for 2 ($\ge 3$) jets}\ \\
\hline \hline  
Signal region & \ 0b-HP\ \ & \ 0b-LP\ \ & \ 1b-HP\ \ & \ 1b-LP\ \ & \ 0b-Res\ \ & 1b-Res \\ \hline 
 $J$ or $jj$&HP & LP & HP &LP & \multicolumn{2}{c}{$\Delta R_{jj} < 1.4$ and $m_{jj} \in [65, 105]$\,GeV} \\
$b$-jet              & no $b$-jet &   no $b$-jet &   1 $b$-jet &   1 $b$-jet &   no $b$-jet &   1 $b$-jet \\ 
\hline\bottomrule 
\end{tabular}
\normalsize
\caption{\label{tab-monoW}
Event selection criteria in the mono $W/Z$ search~\cite{Aaboud:2018xdl}.  
The symbols $j$ and $J$ denote the small-$R$ and large-$R$ jets, respectively. 
$\{j_i\}$ are the small-$R$ jets ordered ($i = 1,2,3,\dots$) by their $p_{\text{T}}$ in decreasing order. 
Angles are defined in radians. See text for details. 
}
\end{table*}
\renewcommand{\arraystretch}{1.0}

\subsection{Mono-boson search selection criteria and $\boldsymbol{E_{\textbf{T}}^{\textbf{miss}}}$ likelihood analysis}
\label{s4.2}

The event selection criteria in the ATLAS mono-$W/Z$ search are given in Table \ref{tab-monoW}. As mentioned above, this search looks for events with large missing transverse energy ($\met$) that contain either a large-$R$ jet (classified as \textsl{merged topology}) or two distinct narrow jets (\textsl{resolved topology}), with dijet invariant mass around that of the $W/Z$ bosons. Jets are clustered according to the anti-$k_t$ algorithm \cite{Cacciari:2008gp} with radius parameter $R=1.0$ (large-$R$) or $R=0.4$ (narrow). In both topologies any events with reconstructed leptons are rejected. In order to suppress multijet backgrounds, the azimuthal separation between the $\met$ vector and the large-$R$ jet is required to be larger than $2\pi/3$ in the merged topology; the same criteria applies in the resolved topology, with the large-$R$ jet replaced by the two-highest-$p_{\text{T}}$-jets system. In addition, the track-based missing transverse momentum $\vec{p}^{\mathrm{miss}}_{\text{T}}$, defined as the negative vector sum of the transverse momenta of tracks with 
$p_{\text{T}} > 0.5$\,GeV and $\abs{\eta} < 2.5$, is required to be larger than 30\,GeV and 
its azimuthal angle to be within $\pi/2$ of that of the calorimeter-based $\met$. In the resolved topology, the highest-$p_{\text{T}}$ jet is required to have $p_{\text{T}} > 45$\,GeV and the sum of $p_{\text{T}}$ of the two (three) leading jets is required to exceed 120 (150)\,GeV.

In addition to the above requirements, in the merged topology any $b$-tagged jet outside the large-$R$ jet is rejected. The signal regions are further classified by the number of $b$-tagged jets (0 or 1) and the \emph{purity} (defined in terms of $p_{\text{T}}$ requrements on the substructure variable $D_2^{(\beta=1)}$ \cite{ATLAS:2015fhd}) of the large-$R$ jet to be tagged as originating from a hadronic vector boson decay. In both signal regions of the resolved topology, the angular separation $\Delta R = \sqrt{(\Delta \phi)^2 + (\Delta \eta)^2}$ between the two leading jets and the invariant mass $m_{jj}$ of the two leading jets is required to be smaller than 1.4 and within a range $[65, 105]$\,GeV, respectively.

In the absence of a discovery, ATLAS imposes limits on an array of BSM scenarios that produce hadronic mono-boson + $E_{\text{T}}^{\text{miss}}$ signals, including exotic invisible Higgs boson decays and vector $Z'$ + dark matter production. An elementary step would be to straightforwardly reinterpret the ATLAS results for electroweakino pairs within our realistic MSSM parameter space, as discussed above. But, as demonstrated in previous work \cite{Carpenter:2021jbd}, we can improve upon a simple recast by exploiting the $E_{\text{T}}^{\text{miss}}$ distributions, which are provided by ATLAS for the observed data and fitted SM background processes.\footnote{The data have been stored by ATLAS on the
\href{https://www.hepdata.net/record/83180}{\textsc{HEPData}} repository (hyperlinked).} The backgrounds considered by ATLAS include $t\bar{t}$ production, SM $W/Z + \text{jets}$ processes (both quite large), and diboson and single-$t$ processes (much smaller). Of these, $W/Z + \text{jets}$ is the dominant background in all signal regions requiring zero $b$-tagged jets --- which are \emph{a priori} most relevant to our electroweakino signals because bottom quarks only appear in $\sim\!\!\!15\%$ of the $\tilde{\chi}\tilde{\chi} + Z$ events, themselves subdominant to $\tilde{\chi}\tilde{\chi} + W^{\pm}$; \emph{viz}. Figure \ref{fig:mono-boson-xsec}.

To execute an analysis based on this ATLAS search, we generate $\tilde{\chi}\tilde{\chi}+V$ events using \textsc{MadGraph5\texttt{\textunderscore}aMC@NLO} version 2.7.2 
and simulate showering and hadronization with {\sc Pythia\,8} version 8.245 \cite{Sjostrand:2014zea}. The signal normalizations are given by the cross sections discussed above. We use {\sc Delphes\,3} version 3.4.2 \cite{deFavereau:2013fsa} as our detector simulator. We modify the default ATLAS \textsc{Delphes} card to include a collection of large-$R$ jets in addition to the standard $R=0.4$ jets. Pile-up is controlled by trimming from large-$R$ jets all $R=0.2$ sub-jets with $p_{\text{T}}$ below 5\% of the original jet $p_{\text{T}}$ \cite{Krohn:2009th}. The energy fractions of chargino tracks in the electromagnetic and hadronic calorimeters are set to zero since, in the model parameter space, the charginos decay too promptly to deposit energy in the calorimeters.  To appropriately capture the physical transition from the higgsino-like region to the well-mixed region, where the neutralino splitting becomes too great for $\tilde{\chi}^0_2$ to be appropriately recorded as missing energy, we veto the production of the second lightest neutralino $\tilde{\chi}^0_2$, at the generator level, wherever $m_{\tilde{\chi}_2^0}-m_{\tilde{\chi}_1^0} > 8\,\text{GeV}$. Finally, since the selection criteria in the analysis~\cite{Aaboud:2018xdl} is adjusted such that the efficiency is $50\%$ independent of jet $p_{\text{T}}$ \cite{ATLAS:2015fhd}, we treat half of the events with a large-$R$ jet as high-purity (HP) events, and the rest are classified into the low-purity (LP) regions. The selections in Table \ref{tab-monoW} are imposed on our event samples, and their efficiencies computed, by an in-house C code used to call the \textsc{ExRootAnalysis} library.

The merged-topology high-purity signal region with zero $b$-tagged jets, 0b-HP, turns out to be most sensitive to our electroweakino signals. This is due in large part to its powerful suppression of the $W/Z + \text{jets}$ backgrounds mentioned above. The 0b-HP selection is effective at cutting away these backgrounds because their missing energy is generated by leptonically decaying vector bosons, hence --- for events passing the stringent $E_{\text{T}}^{\text{miss}} > 250\,\text{GeV}$ selection --- the large-$R$ jet requirement in the high-purity region can only be satisfied by accidental reconstruction from the QCD multijet background. Since we have found consistently, beginning with even earlier work \cite{Carpenter:2020fnh}, that the 0b-HP signal region gives the strongest bounds, we focus on this region in what follows.

We now return to the $E_{\text{T}}^{\text{miss}}$ distributions, which are our point of departure from the original ATLAS analysis. In Figure 1 of our previous work \cite{Carpenter:2021jbd}, for illustrative purposes, we compared the $E_{\text{T}}^{\text{miss}}$ distributions in the 0b-HP signal region for data and SM background to the $\tilde{\chi}\tilde{\chi}+V$ signal in two higgsino-like LSP scenarios with $\mu = 200\,\text{GeV}$ and $\mu = 500\,\text{GeV}$. The missing energy recorded in 0b-HP events is divided into eight bins of increasing width between 200 and 1500\,GeV, with the last bin $E_{\text{T}}^{\text{miss}} \in [800,1500]\,\text{GeV}$. To obtain the binned yields for those signals, additional selections corresponding to these $E_{\text{T}}^{\text{miss}}$ bins were added to our analysis code at that time. Crucially, we found that the background $E_{\text{T}}^{\text{miss}}$ falls more quickly than that of the higgsino signals. We now find similar behavior in wino-like LSP scenarios. This implies that more stringent cuts on $E_{\text{T}}^{\text{miss}}$ may produce improved sensitivity to progressively heavier electroweakinos throughout the $(\mu,M_2)$ space with suitable mass splitting(s).

For this work, with the yields computed (including the $E_{\text{T}}^{\text{miss}}$ binning) for our signals throughout the $(\mu,M_2)$ plane, we perform a joint-likelihood analysis assuming Poisson-distributed data and Gaussian backgrounds such that the likelihood function takes the form \cite{ATLAS:2011tau}
\begin{multline}\label{eq:likelihood}
\mathcal{L}(m \mid \mu,b) = \prod_{i=1}^{N_{\text{bin}}} \frac{(\mu s_i + b_i)^{m_i}}{m_i!}\, \text{e}^{-(\mu s_i + b_i)}\\ \times \frac{1}{\sqrt{2\pi}\,\sigma_{b,i}} \exp \left\lbrace -\frac{1}{2}\frac{(b_i-\langle b_i\rangle)^2}{\sigma_{b,i}^2}\right\rbrace.
\end{multline}
The yield (data) in each bin $i$ is $m_i$. The signal yield according to an alternate hypothesis is $s_i$ with strength modifier $\mu$. The background distribution in each bin is centered at $\langle b_i \rangle$ and has uncertainty $\sigma_{b,i}$. We use the joint likelihood to compute the test statistic
\begin{align}\label{eq:testStatistic}
q_{\mu}^m = -2 \ln \frac{\mathcal{L}(m \mid \mu,\hat{\hat{b}})}{\mathcal{L}(m \mid \hat{\mu},\hat{b})},\ \ \ \hat{\mu} \leq \mu,
\end{align}
where $\hat{\hat{b}} = \hat{\hat{b}}(\mu)$ in Eq.~\eqref{eq:testStatistic} is the conditional maximum-likelihood (ML) estimator of the likelihood for a given $\mu$ and the pair $(\hat{\mu},\hat{b})$ are the unconditional ML estimators \cite{Cowan:2010js}. The one-sided limit at 95$\%$ C.L. is then given in terms of \eqref{eq:testStatistic} by
\begin{align}
\text{CL}_s = 0.05 = \frac{1- \Phi([q_{\mu=1}^{m=n_{\text{obs}}}]^{1/2})}{\Phi([q_{\mu=1}^{m=\langle b\rangle}]^{1/2}-[q_{\mu=1}^{m=n_{\text{obs}}}]^{1/2})},
\end{align}
where $\Phi$ is the cumulative distribution function of the normal distribution with zero mean and unit variance and $n_{\text{obs}}$ is the true number of events surviving the experimental selection \cite{ATLAS:2020yaz}. In addition to computing the sensitivity of the search given the real data, we make rough sensitivity projections for HL-LHC by rescaling the yields by a factor $\mathcal{R}(\mathcal{L}) = \mathcal{L}/(36.1\,\text{fb}^{-1})$ and the (background) uncertainties by a factor $\sqrt{\mathcal{R}(\mathcal{L})}$. We then compute the median significance for exclusion and discovery of our signal according to
\begin{align}\label{eq:exclusionDiscovery}
Z_{\text{excl}} \equiv [q_{\mu=1}^{m=\langle b \rangle}]^{1/2}\ \ \ \text{and}\ \ \ Z_{\text{disc}} \equiv [q_{\mu=0}^{m=s + \langle b \rangle}]^{1/2},
\end{align}
taking $Z_{\text{excl}} = 2$ and $Z_{\text{disc}} = 5$ as our exclusion and discovery thresholds.

\section{Results}\label{s5}

\begin{figure*}
    \centering
    \includegraphics[width=0.95\linewidth]{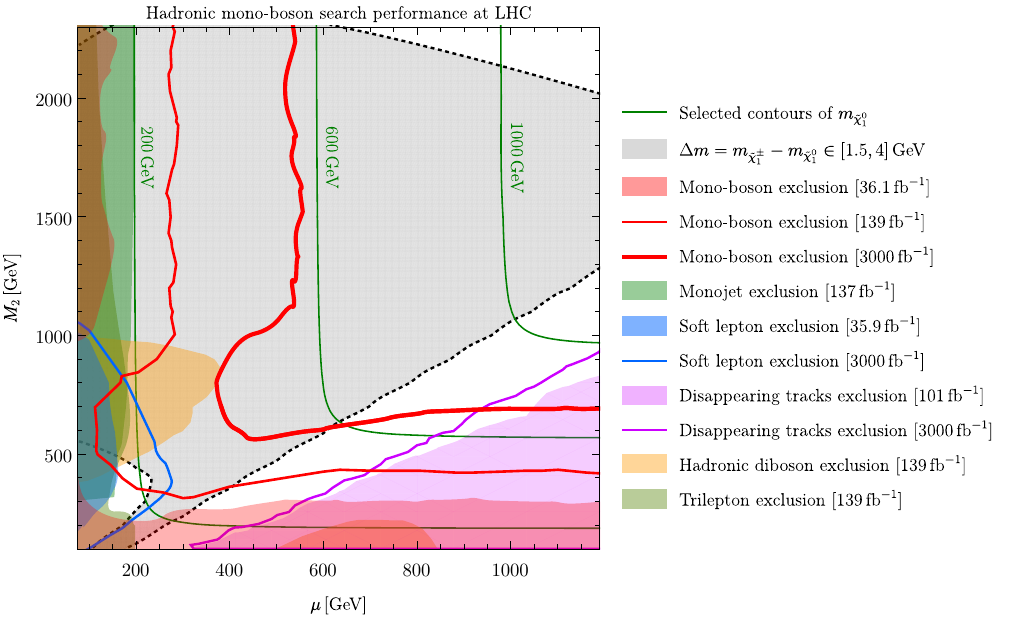}
    \caption{\label{fig:limits_plot} Performance projections for custom hadronic mono-$W$/$Z$ search for the original $36.1\,\text{fb}^{-1}$ dataset (red shaded) and for the full Run 2 and HL-LHC datasets (red curves) compared to existing searches for electroweakinos decaying to two soft leptons (CMS-SUS-16-048, blue) and with tracks disappearing in the silicon tracker (CMS-EXO-19-010, violet). Also included are conventional searches for electroweakino pair production in final states with two hard hadronic vector bosons (ATLAS-SUSY-2018-41, orange) and three leptons (ATLAS-SUSY-2019-09, light green), along with the strongest monojet + $E_{\text{T}}^{\text{miss}}$ search (CMS-EXO-20-004, dark green).
    }
\end{figure*}

We now present results as exclusions in the $(\mu,M_2)$ parameter space. Following our work in reference \cite{Carpenter:2021jbd} we determine the statistical significance of the mono-boson analysis by constructing a joint-likelihood function from our binned missing energy analysis within the 0b-HP signal region. The limits from this search and several others are displayed in Figure \ref{fig:limits_plot}. In this figure the green contour lines show a few distinct lightest neutralino masses $m_{\tilde{\chi}^0_1}$. There is a shaded region in the background in which the mass difference between the lightest neutralino $\tilde{\chi}_1^0$ and the lightest chargino $\tilde{\chi}_1^{\pm}$, first seen in Figure \ref{fig:N1C1_mass_difference_plane}, is between $1.5\,\text{GeV}$ and $4.0\,\text{GeV}$. The red shaded region indicates limits from the mono-boson search at 95$\%$ C.L. in the mass plane for the original $\mathcal{L}=36.1\,\text{fb}^{-1}$ dataset, while the thin and thick red contours represent exclusion projections for (respectively) the full Run 2 dataset of $\mathcal{L}=139\,\text{fb}^{-1}$ and the HL-LHC with $\sqrt{s}=14\,\text{TeV}$ and $\mathcal{L} = 3\,\text{ab}^{-1}$. 

In the upper region of this plot, for higgsino-like LSPs, the projected sensitivity hovers around 150\,GeV for the Run 2 LHC dataset with integrated luminosity $\mathcal{L} = 139\,\text{fb}^{-1}$ and is anticipated to reach over 550\,GeV for the HL-LHC run with $\mathcal{L} = 3\,\text{ab}^{-1}$. The $5\sigma$ discovery sensitivity for HL-LHC, which was calculated but is omitted from the plot for visual clarity, is around 300\,GeV. As long as $M_2$ is sufficiently above $\mu$, the lightest neutralino mass --- and therefore the lower mass bound --- is relatively independent of $M_2$, since the neutralino maintains a sufficiently higgsino-like admixture. In the mixed wino-higgsino region, we project that the $139\,\text{fb}^{-1}$ dataset has exclusion sensitivity up to $m_{\tilde{\chi}^0_1} \sim 200\,\text{GeV}$. The HL-LHC exclusion sensitivity reaches past 600\,GeV for these well-mixed states, with $5\sigma$ discovery sensitivity at around 450\,GeV. It is evident that the limit strengthens both in the pure higgsino region and in the well-mixed region, with a noticeable dip between these regions. This can be explained by considering that for fixed $\mu$, as $M_2$ decreases, the mass difference between $\tilde{\chi}^0_2$ and $\tilde{\chi}^0_1$ (not shown here, but plotted in Figure \ref{fig:N1N2_difference}) increases such that $\tilde{\chi}^0_2$ decay through off-shell gauge bosons with appreciably hard decay products no longer appear as invisible particles contributing to the search, while the corresponding decay through an on-shell vector boson illustrated in Figure \ref{fig:channel_diagrams} has not yet ``turned on'' sufficiently to contribute to the $V + E_{\text{T}}^{\text{miss}}$ channel.

As mentioned in \hyperref[s3]{Section III}, we wish to compare the sensitivity of the mono-boson search to long-lived track and soft-lepton searches that may be more powerful in parameter space with different electroweakino spectra. Limits from CMS-EXO-19-010 and CMS-SUS-16-048 are therefore included in Figure \ref{fig:limits_plot} as violet and blue regions/contours, respectively. These shaded regions, and all of their counterparts discussed below, denote observed limits. In analogy with the mono-boson search, shaded regions indicate current limits at 95\% C.L. and solid curves represent HL-LHC projections computed using \textsc{MadAnalysis\,5} (\emph{viz}. \hyperref[s3]{Section III}). The logic discussed in that section is borne out in Figure \ref{fig:limits_plot}: the two CMS searches constrain parameter space complementary to that probed by the mono-boson search. In particular, the mono-boson sensitivity gap between the higgsino-like and well-mixed regions, discussed just above, is filled to some extent by the soft dilepton analysis. Meanwhile, the long-lived track search is several times more powerful than the mono-boson analysis (as a function of $M_2$) in the wino-like region. It is worth noting that these searches cannot match the HL-LHC gains of the mono-boson search for the higgsino region, $\mu \ll M_2$, on the basis of improved statistics, simply because their sensitivities are naturally limited to parameter space with suitable electroweakino mass splitting(s). This is also true of the long-lived/disappearing-track search in the mixed wino-higgsino region, which is only sensitive to charginos with lifetimes exceeding $\tau = 0.3\,\text{ns}$ --- and are already well constrained with Run 2 data.

We next return to monojet constraints, first mentioned in \hyperref[s3.3]{Section III}. Shaded in green on the left edge of Figure \ref{fig:limits_plot} are the strongest available monojet limits, which come from the $137\,\text{fb}^{-1}$ monojet subanalysis of CMS-EXO-20-004 \cite{CMS:2021far}. These limits were very recently calculated for simplified pure-higgsino (LSP) parameter space in \cite{Agin:2023yoq} (including one of the authors of this work) using the implementation of this analysis in the \textsc{MadAnalysis\,5} PAD and the statistical analysis package \textsc{Spey} \cite{Araz:2023bwx}. We focus on the CMS limits since they are much stronger than the available recast ATLAS limits: this is because CMS has published the correlations between signal regions for the background model in a simplified-likelihood framework, permitting the computation of a limit based on the signal region combination, whereas ATLAS provides no statistical information in \cite{ATLAS:2021kxv} and the best limit comes only from the most sensitive individual signal region. For this work, we have mapped the CMS results onto the $\mu \ll M_2$ region of our parameter space, where the simplified pure-higgsino model provides a good approximation to the true mass spectrum. A similar analysis has yet to be carried out for pure-wino LSP models, but by comparison with the higgsino limits we expect monojet limits on winos to be superseded by disappearing-track bounds in most of the wino-like region. The higgsino monojet limits weaken rapidly as the splitting between light neutralinos $m_{\tilde{\chi}^0_2}-m_{\tilde{\chi}^0_1}$ approaches $20\,\text{GeV}$ due to vetoes on leptons with $p_{\text{T}} > 10\,\text{GeV}$, which can result from off-shell $W/Z$ bosons in electroweakino decays. Ultimately, we find that the $137\,\text{fb}^{-1}$ CMS monojet (observed) limits are stronger than the ``true'' $36.1\,\text{fb}^{-1}$ mono-boson limits (recall these are shaded in red), excluding up to $\mu \approx 200\,\text{GeV}$. Our projection shows that the improved mono-boson analysis takes back the lead when the yields are rescaled to $139\,\text{fb}^{-1}$ to estimate the full Run 2 sensitivity. We therefore conclude that the mono-boson analysis remains superior to monojet searches --- at least for higgsinos, for which these analyses compete to set the best limits --- when the datasets are of approximately equal size.

Finally, as alluded to in \hyperref[s4]{Section IV}, we demonstrate the complementarity between the searches detailed above, which explicitly target compressed spectra, and more conventional searches for electroweakino pair production. In light green we represent the observed limits from another $139\,\text{fb}^{-1}$ ATLAS search, ATLAS-SUSY-2019-09 \cite{ATLAS:2021moa}, which combines a search for final states with three leptons and missing transverse momentum with a previous $139\,\text{fb}^{-1}$ search for soft-dilepton + $E_{\text{T}}^{\text{miss}}$ final states \cite{ATLAS:2019lng}. (This soft-dilepton analysis constitutes a significant update to the $35.9\,\text{fb}^{-1}$ CMS soft-lepton search discussed above.) As explained in the previous section, this search topology results from soft decays of the chargino and second lightest neutralino. One of the scenarios considered by ATLAS contains compressed higgsino-like electroweakinos, so in the absence of a dedicated recast we perform a simple mapping from the physical plane ($m_{\tilde{\chi}^0_2},m_{\tilde{\chi}^0_2} - m_{\tilde{\chi}^0_1})$ presented by ATLAS onto our $(\mu,M_2)$ plane, using our spectra computed by \textsc{SPheno} as discussed in \hyperref[s2]{Section II}. Exclusions from this search overlap with exclusions from the soft-lepton search and fade out as we enter the higgsino-like LSP region, where decay products become invisible; and as we approach the well-mixed region, where heavier neutralino and chargino production is mass suppressed. The mono-boson search dominates the trilepton exclusions for sizable $\mu$, and presumably a combination of these search channels would increase constraints where the two searches are roughly equally powerful. Moving on, in orange we denote the space excluded by ATLAS-SUSY-2018-41, a search for pair-produced electroweakinos with two hadronically decaying vector bosons and missing energy \cite{ATLAS:2021yqv}. This search uses the full Run 2 dataset of $\mathcal{L} = 139\,\text{fb}^{-1}$ and uniquely (among the analyses discussed in this work) presents results in the $(\mu,M_2)$ plane that can be included without further comment. It relies on the production and decay of heavy $\tilde{\chi}_2^{\pm}$ and $\tilde{\chi}_3^{0}$ states and has exclusion power where they are light enough to have sufficient production cross sections but heavy enough to produce a vector boson hard enough for a boosted tag upon decay. In the deep wino and higgsino regions, these electroweakinos are too heavy for sufficient production rates.

Altogether, we find that the mono-boson search should do the heavy lifting in the $(\mu,M_2)$ plane during the run of HL-LHC. Nevertheless, analyses of all types have a role to play in probing this space, with long-lived tracks searches covering the deep wino region, our mono-boson analysis offering excellent constraints through a wide coverage of the plane, and \emph{e.g.} the hadronic diboson search filling gaps in the mono-boson analysis as $\mu$ begins to approach $M_2$ from above. Taken together, if no excess is measured, we project that these  analyses can exclude much of the parameter space with $\mu \lesssim 500\,\text{GeV}$ and $M_2 \lesssim 500\text{--}750\,\text{GeV}$ by the end of the LHC's high-luminosity run. This is of some interest, as the size of the $\mu$ parameter itself has long been proposed as a measure of the electroweak fine-tuning of supersymmetric scenarios as given by the minimization conditions of the Higgs potential \cite{Chan:1997bi,Baer:2012up}.  This measure of naturalness requires the $\mu$ term not exceed a few hundred GeV, so by this metric we find that our HL-LHC search will have the power to exclude the ``natural'' region of the MSSM.

\section{Conclusions}\label{s6}
In this work we have explored multiple experimental handles on the relatively unconstrained wino-higgsino plane ($\mu,M_2$). We have proposed a hadronic mono-boson search with binned $E_{\text{T}}^{\text{miss}}$ selections as an LHC channel sensitive to neutralinos with significant higgsino admixtures. We have reviewed how the light electroweakino states vary in mass and content in the $(\mu,M_2)$ plane and described how production processes relevant to the mono-boson search depend on the mass splittings between the $\tilde{\chi}_1^0, \tilde{\chi}^{\pm}_1$ and $\tilde{\chi}_2^0$ states. We have also highlighted other search strategies --- targeting events with soft leptons, events with long-lived tracks that disappear before the edge of the tracker, and searches with moderately heavy but producible electroweakino pairs --- that constrain wino-higgsino parameter space complementary to that probed by the mono-boson search.

We have set limits based on our proposed strategy and from reinterpreted existing results using LHC Run 2 data, and we have performed a sensitivity study for the $3\,\text{ab}^{-1}$ run of the HL-LHC. We have depicted these limits in a considerable portion of the $(\mu,M_2)$ plane. If no excess is seen, the mono-boson search has sensitivity to pure- or nearly-pure-higgsino LSPs of mass $m_{\tilde{\chi}^0_1}\sim 150\,\text{GeV}$ and mixed wino-higgsino LSPs up to 300 GeV in the current data set. It also has the power to exclude almost all $M_2 < 1\,\text{TeV}$ for $\mu \sim 120\,\text{GeV}$ and all $\mu < 1\,\text{TeV}$ for $M_2 \sim 250\,\text{GeV}$ when combined with other recast search limits. At the HL-LHC, for $M_2 \sim 750\,\text{GeV}$, we project a lower bound of $\mu\approx 400\,\text{GeV}$ in the entire mass plane with exclusions (assuming no excess is observed) of higgsino-like neutralinos up to 550\,GeV, and past 600\,GeV in the well-mixed region. We also project $5\sigma$ discovery potential up to 300\,GeV for a higgsino-like LSP and up to 450\,GeV for mixed wino-higgsino LSP.  

As hoped, we have found that the soft-lepton, disappearing-track, and boosted diboson searches are sensitive to $(\mu,M_2)$ parameter space in which the mono-boson analysis is weak, thus exhibiting useful complementarity. Exclusions from events with soft but detectable leptons and the diboson analysis fill a notable gap in the mono-boson analysis for low-mass LSPs between the higgsino-like and well-mixed regions, while wino-like long-lived charginos with $M_2 \lesssim 1\,\text{TeV}$ are most strongly constrained by the disappearing-track search. We project that this complementarity will allow the HL-LHC to rule out vast swaths of ``natural'' (sub-TeV) wino-higgsino parameter space in the absence of a discovery.

We expect these results to be somewhat robust with respect to the bino mass $M_1$, which we mentioned in \hyperref[s2]{Section II} was decoupled in our analysis. We know, for instance, that the specific choice of $M_1=5\,\text{TeV}$ can be relaxed to as low as $2\,\text{TeV}$ with negligible effect on the electroweakino spectrum. The exclusions in Figure \ref{fig:limits_plot} will quantitatively change if $M_1$ is taken much lower, in the vicinity of $\mu$ or $M_2$ (whichever is heavier), but the picture will remain qualitatively the same --- including which searches are most sensitive in general regions of the $(\mu,M_2)$ plane --- as long as the bino is still heavier than the wino-higgsinos. Only when the bino is lighter than one or both of the higgsino or wino will the results cease to apply even qualitatively, so that a new (meta-)analysis will be required. Since scenarios with light binos naturally produce larger electroweakino mass splittings, we expect conventional searches, including for instance the trilepton search discussed in \hyperref[s5]{Section V}, to dominate searches targeting compressed spectra and exclude much more parameter space. But we reiterate that an accurate and comprehensive picture can be painted in some future project analogous to the present work.

Even within the decoupled-bino paradigm discussed in this work, opportunities for further study are numerous. There may be opportunities for the study of mono-boson signatures of electroweakinos in which the boson decays leptonically. A search for a single leptonically decaying mass-reconstructed $Z$ boson was previously proposed for dark matter and higgsino LSPs \cite{Anandakrishnan:2014exa,Carpenter:2012rg}, and such strategies might be applied to the entire $(\mu,M_2)$ plane. Leptonic mono-$W$ searches with a leptonic transverse-mass cut and a binned missing energy analysis might also provide a probe of the wino-higgsino plane. Such leptonic analyses might be interesting in light of the current excess \cite{ATLAS:2019lng,Agin:2023yoq} in events with soft lepton pairs. Finally, combinations of the analyses in this work could provide tighter constraints on the $(\mu,M_2)$ plane for the existing LHC dataset.

\acknowledgments

The work of L. M. C., H. B. G., and T. M. was supported in part by the Department of Physics at The Ohio State University. L. M. C. is further supported by Grant DE-SC0024179 from the United States Department of Energy (DOE). H. B. G. is further supported by an MPS-Ascend Fellowship, Award 2213126, from the United States National Science Foundation (NSF). J. K. is supported in part by Grant IBS-R018-D1 from the Institute for Basic Science (IBS), Korea. T. M. is further supported by Grant ANR-21-CE31-0013, Project DMwithLLPatLHC, from the \emph{Agence Nationale de la Recherche} (ANR), France.

\bibliographystyle{apsrev4-2}
\bibliography{reference}

\end{document}